\documentclass[aps,prl,preprint,superscriptaddress,showpacs]{revtex4}
\usepackage{amssymb}
\usepackage{graphicx}
\newcommand{\ud}{\mathrm{d}}
\newcommand{\expo}{\mathrm{exp}}

\begin{document}


\title{Persistent spin currents in electron systems with spin-orbit interaction}


\author{Xiang Zhou}
\email[To whom correspondence should be addressed:~]{xzhou@mac.com}
\affiliation{Department of Physics, Wuhan University, Wuhan 430072, China}
\affiliation{Key Laboratory of Acoustic and Photonic Materials and Devices of Ministry of Education, Wuhan 430072, China}
\author{Cheng-Zheng Hu}
\affiliation{Department of Physics, Wuhan University, Wuhan 430072, China}
\affiliation{Key Laboratory of Acoustic and Photonic Materials and Devices of Ministry of Education, Wuhan 430072, China}
\author{Zhenyu Zhang}
\affiliation{Department of Physics, Wuhan University, Wuhan 430072, China}
\affiliation{Key Laboratory of Acoustic and Photonic Materials and Devices of Ministry of Education, Wuhan 430072, China}
\author{Ling Miao}
\author{Xia Wang}
\affiliation{Department of Physics, Wuhan University, Wuhan 430072, China}


\date{\today}

\begin{abstract}
We investigate the persistent spin currents in one- and two-dimensional electron systems with spin-orbit interaction in thermodynamics equilibrium at absolute zero temperature. The persistent spin current is the intrinsic one which is connected with the Berry phases in the configuration spaces of an electron system and winding numbers in the field configurations of electrons.  When the topological space of the configuration of a system has the nontrivial first homotopy groups, the persistent spin currents in the system could be nonzero and not easily destroyed by impurity scattering in ballistic limit.  The non-vanishing background spin currents in infinite two-dimensional electron system found by Rashba could be realized by the transport persistent spin currents in a finite torus electron system with spin-orbit interaction.  In this sense, we meet the challenge proposed by Rashba.

\end{abstract}


\maketitle

Intrinsic spin current (SC) could be generated in two-dimensional electron system (2DES) with spin-orbital interaction (SOI) by universal intrinsic spin Hall effect (SHE)\cite{Sinova04}. It is connected to the Berry phase in the momentum space of the 2DES and winding numbers (WNs) of the spin-orbit field\cite{Raichev07}.  The experimental observations of SHE in 2DES have been reported\cite{Awschalom}.  However, whether the electron SHE is intrinsic or not is still under debate because the intrinsic SCs would be destroyed by impurity scattering even in the ballistic regime\cite{non}.  Besides, the existence of intrinsic SCs faces a challenge proposed by Rashba\cite{Rashba03}.  He found that there are nonzero intrinsic SCs in an infinite 2DES with SOI in the thermodynamic equilibrium.  In his opinion, the background SCs are the non-transport ones and should be eliminated in calculating the transport SC by modifying the definition of the SC. After his work, many subsequent studies have discussed the definition of the SC\cite{scdef}.  There is not a satisfied explanation yet.

The standard Hamiltonian of 2DES with Rashba SOI is given by
\begin{equation}
  \label{eq:1}
  H=\frac{1}{2}p^2 - \lambda\vec{\sigma}\cdot(\hat{z}\times\vec{p}),
\end{equation}
where $\lambda$ is the Rashba coupling constant, $\vec{\sigma}$ is the Pauli matrices and $\hat{z}$ is the unit vector perpendicular to the confinement plane.  For notational simplicity the electron effective mass $m$ and $\hbar$ are set to unity .  The Rashba SOI, which is linear momentum-dependent, provides an effective gauge potential $\Lambda=\lambda\epsilon_{ij3}$, an analog to the electromagnetic vector potential $\mathbf{A}$, where $\epsilon_{ij3}$ is the antisymmetric tensor of which the third subscript correspond to the $\hat{z}$ direction.  There is a coupling of the spin current $\mathbf{j}^s$ and the effective gauge potential $\Lambda$ which would induce the Aharonov-Casher (AC) effect, one of realizations of the Berry phase in the presence of SOI\cite{Aharonov84,Balatsky93}.  It was found about two decades ago that the persistent SCs could be caused by the AC effect in the absence of external magnetic flux in mesoscopic rings\cite{Balatsky93,Loss90}.   Recently, it was predicted that a persistent SC could be realized in the nonmagnetic semiconductor ring\cite{Sun07}.

In this letter, we study the relationship between the persistent SCs and the topological properties of the configurations of systems in one-dimensional (1D) and two-dimensional (2D) mesoscopic electron systems with SOI in thermodynamic equilibrium at absolute zero temperature.  The existence, a main property, of the persistence SCs could be gained without solving the Schr\"odinger equation.

\begin{figure}[!hbp]
\includegraphics{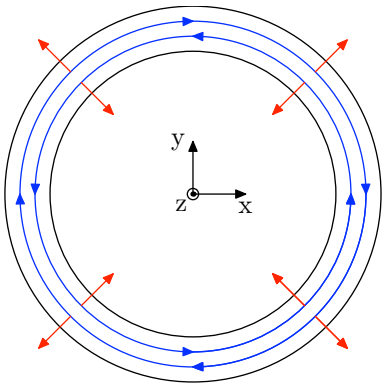}
\caption{Schematic diagram for the circular movement (labeled by blue arrows) of polarized spin (labeled by red arrows) in a ring.\label{ring}}
\end{figure}

Usually the exact analytical results can be obtained in 1D systems where the topological properties would be easily explored.  We first consider a mesoscopic ring fabricated from a 2DES system with Rashba interaction in which only the lowest transverse eigenstate is populated.  The ring system could be treated as a strictly one-dimensional electron system (1DES).  For notational simplicity, the radius of the ring $R$ is set to unity.  The  Hamiltonian of the system is
\begin{equation}
  \label{eq:1dh}
  H=-\frac{1}{2}\frac{\partial^2}{\partial\theta^2}+i\lambda\sigma_r\frac{\partial}{\partial\theta},
\end{equation}
where the $\sigma_r=\vec{\sigma}\cdot\hat{r}$ is the Pauli matrix in radial direction.  The periodic boundary condition is $\psi(0)=\psi(2\pi)$.  There is a circular spin-polarized current in the ring, see Fig.~\ref{ring}.

To explore the relationship between the persistent SC and the topology of ring configuration, we formulate the system in the language of the Euclidean path integral\cite{Justin96,Altland06}.  The quantum partition function (QPF) is
\begin{eqnarray}
  \label{eq:epi}
  \mathcal{Z} & = & \int_0^{2\pi}\ud\theta<\theta|e^{-\beta H}|\theta> \\\nonumber
  & = & \int_0^{2\pi}\ud\theta\int_{\theta(\beta)=\theta(0)}D\theta(\tau)\expo[-\int_{0}^{\beta}\ud\tau L(\theta,\dot{\theta})],
\end{eqnarray}
where $\beta=\frac{1}{kT}$ and the image time Lagrangian is given by
\begin{equation}
  \label{eq:il}
  L(\theta,\dot{\theta})=\frac{1}{2}\dot{\theta}-i\sigma_{r}\lambda\dot{\theta}+\frac{1}{2}\lambda^2.
\end{equation}
The trace implies that path $\theta(\tau)$ must start and finish at the same point. The periodic condition accommodate the invariance of the field configuration $\theta$ under translation by $2\pi$.  The QPF can be cast in the form
\begin{widetext}
%
\begin{equation}
  \label{eq:qpf}
\mathcal{Z}= \sum_{\sigma}\int_0^{2\pi}\ud\theta\sum_{n=-\infty}^{\infty}\int_{\theta(\beta)=\theta(0)+2\pi n}\!\!\!\!\!e^{2\pi in\sigma\lambda}D\theta(\tau)\expo[-\int_0^{\beta}\ud\tau\,(\frac{1}{2}\dot{\theta}^2+\frac{1}{2}\lambda^2)]= \sum_{\sigma} \sum_{n=-\infty}^{n=\infty}e^{-\beta(\frac{1}{2}n^2-n\sigma\lambda)},
\end{equation}
\end{widetext}
where $\sigma=\pm 1$ is the eigenvalue of $\sigma_r$ and $n$, a winding number (WN) of field configuration\cite{tpcomment}, is the number of times $\theta(\tau)$ winds around the unit ring as image time $\tau$ progresses from $0$ to $\beta$.  The periodic boundary condition makes the field configuration $\theta(\tau)$ a covering map $\mathbb{R}\to S^1$ from the interval $[0,\beta]$ into a circle.  The map is classified by the first homotopy group $\pi_{1}$ of the topological space $S^1$ of the ring.  The group $\pi_{1}(\mathrm{S}^1)$ is isomorphic to $\mathbb{Z}$ where the elements are called the WNs\cite{Nakahara90,Munkres00,Fulton95}.

\begin{figure}[!hbp]
\includegraphics{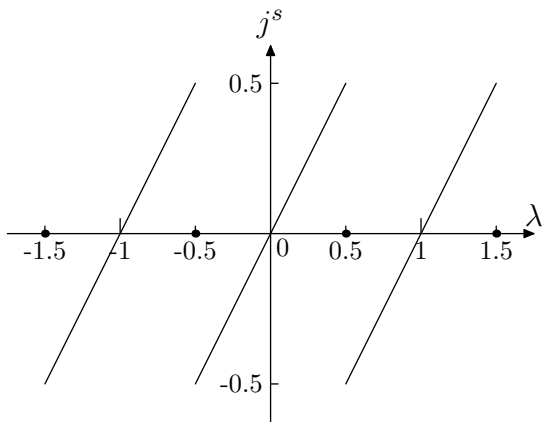}%
\caption{\label{sc} The values of the circular total persistent SC in a ring where $\lambda$ in units of $\hbar^2/mR$ and $j^s$ in unites of $N\lambda$.}
\end{figure}

One can get the spectrum of system
\begin{equation}
  \label{eq:eigen}
  \epsilon_{n,\pm}=\frac{1}{2}n^2\mp n\lambda, \quad n\in\mathbb{Z}.
\end{equation}
According to the conventional definition of the SC, the expectation value of the persistent SC given by every eigenstate is
\begin{equation}
  \label{eq:eigensc}
  j^{s}_{n,\pm}=<\psi_{n,\pm}|\frac{1}{4}\{\sigma_r,v\}|\psi_{n,\pm}>=\frac{1}{2}\frac{\partial\epsilon_{n,\pm}}{\partial n}.
\end{equation}
The total persistent SC in the ring is $j^s=\sum j^s_{n,\pm}$, where the summation should be performed over all $N$ eigenstates under the Fermi energy.  The total persistent SC of the ring is a sawtooth function of $\lambda$, plotted in Fig.~\ref{sc}.  When $\lambda$ is an integer or half-integer, the total persistent SC is zero because the persistent SC given by the eigenstates under the Fermi energy cancel each other.  The persistent SC are continuous to zero in the neighborhood of the integer $\lambda$ while they discontinuous to zero and oscillate between $\frac{1}{2}N\lambda$ and $-\frac{1}{2}N\lambda$ in the neighborhood of the half-integer $\lambda$.  Our results qualitatively accord with others obtained by canonical quantization method\cite{Sun07}.  Since all simple closed loop systems are homeomorphic to each other, by the above Euclidean path integral method one can readily find that the persistent SCs have the same properties if both simple closed loop systems with SOI have the same circumference.

There is a $\lambda$-dependent term $S_{\mathrm{top}}$ of the image time action in the QPF,
\begin{equation}
  \label{eq:stop}
  S_{\mathrm{top}}\equiv i\sigma\int_0^{\beta}{\rm{d}}\tau\,\lambda\dot{\theta}=\sigma\lambda(\theta(\beta)-\theta(0))=\sigma i2\pi n\lambda,
\end{equation}
which involves the WN of field configuration.  The exponentiation of $S_{\mathrm{top}}$ contributes a complex number to QPF.  Eq.(\ref{eq:stop}) makes the topological aspects of the ring system particularly transparent.  The integer and half-integer $\lambda$s, of which the total persistent SC is zero, give the exponentiation of $S_{\mathrm{top}}$ the real number $+1$ and $-1$, respectively.

The system is gauge invariant under the gauge transformation $\psi(\theta)\to e^{i\Delta\theta}\psi(\theta)$ while the boundary condition is changed to $\psi(0)=e^{i2\pi\Delta}\psi(2\pi)$ and derivative operator $\partial_{\theta}=\frac{\partial}{\partial\theta}$ to covariant derivative operator $D_{\theta}=\partial_{\theta}+i\Delta$.  The presence of the SOI amounts to a twist in the boundary conditions and the persistent SC is a measure of the sensitivity of the spectrum to the twist\cite{Altland06}.  In ballistic limit the impurity scattering would not easily destroy the coherence of wave functions represented by the WNs of the field configuration.  The effect of the impurity scattering would be resist by the discontinuousness of the WNs of the field configurations.

The periodic boundary condition is a restriction to the gauge potential.  If we keep the boundary condition periodic, $\Delta$ must be an integer and the gauge transformation is discrete.  In the ring system, the effective gauge potential in Hamiltonian is reduced to a scalar $\lambda$.  When $\lambda$ are integers or half-integers, the system has the extra symmetry corresponding to discrete gauge transformation and then the total persistent SC must be zero.

When the topology of the configuration of a system is changed, the persistent SCs in it would be changed extremely.  If a thin radial cut is made in a ring, its topological space denoted $X$ is trivially homeomorphic to a closed interval and the first homotopy group $\pi_1(X)$ is the trivial (one-element) group.  The boundary condition is $\psi(0)=\psi(2\pi)=0$ which is confined and lacks the periodicity.  The spectrum of the cut ring system is
\begin{equation}
  \label{eq:eigenrc}
  \epsilon_{n,\pm} = \frac{1}{8}n^2-\frac{1}{2}\lambda^2, \quad n\in\mathbb{Z}.
\end{equation}
Since the spectrum of the new system is symmetric to $n$ whatever $\lambda$ is, there is no persistent SC in it.  If it were nonzero, otherwise there would be infinite accumulation of spin at boundaries.  

It has been shown that the persistent SC is connected with Berry phase in configuration space and WNs of field configuration which is connected with the first homotopy group, a main concept in algebraic topology, of the configuration of a system.  The existence of the persistent SC is only connected with the property of the first homotopy group.   Thus one can prejudge the existence of the persistent SC without solving the Schr\"odinger equations.  It is convenient to use the topology-theoretical method in 2DES where the solution of the Schr\"odinger equation could not be easily obtained\cite{comonstf}.  Whether the boundary condition in 1DES is periodic or not corresponds to the nontrivial or trivial homotopy group, respectively.  In general, the configuration space of a system is an $n$-dimensional manifold denoted $M$ and the boundary of $M$ denoted $\partial M$ has $(n-1)$ dimensions.  If the manifold is bounded with a boundary, the confined boundary condition could be $\psi|_{\partial M}=0$ if we ignore the edge state of systems.

The manifolds of finite 2D systems are classified by simple connectedness.  The 2D manifold has a trivial homotopy group if it is simple connected where any two paths having the same initial and final points are path homotopic.  Usually, $\psi|_{\partial M}=0$ gives the complete confined boundary condition.  For example, the boundary condition in the unit square SOI system are $ \psi|_{y=0}=\psi|_{y=1}=0,\quad \psi|_{x=0}=\psi|_{x=1}=0$.  One can readily conclude there is no persistent SC, otherwise there would be infinite accumulation of spin at the edge.  However, there is an exception that a persistent SC could exist in a 2D disk with SOI.  The boundary condition $\psi|_{\partial M}=0$ is only confined in the radial direction $\psi|_{r=1}=0$, where we set the radius unit for simplicity.  The angular boundary condition is periodic $\psi(\theta)=\psi(\theta+2\pi)$.  A circular persistent SCs around the center point $O$ could exist in the disk.  In topological viewpoint, the center point $O$ of a 2D disk $D$ is a fixed point under rotation which does not contribute to the circular persistent SC.  Thus the manifold which contribute to the persistent SC is homeomorphic to $D-\{O\}$ where the WNs can be defined. 

\begin{figure}[!hbp]
\centering
\includegraphics{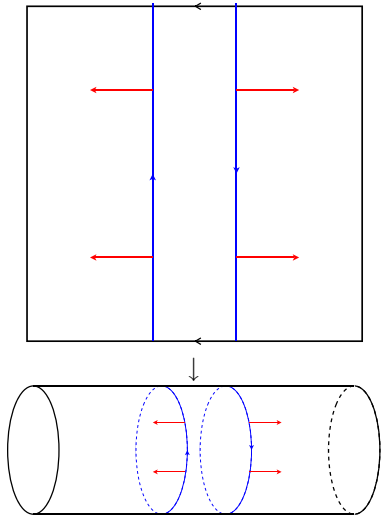}
\caption{Schematic diagram for the circular movement of spin in a cylinder. \label{cylinder}}
\end{figure}

If the manifolds are not simple connected, there must be at least two paths having the same initial and final points which are not path homotopic.  Thus there are some non-trivial topological terms of the image time action in QPF which would affect the spectrum of the systems and give the nonzero persistent SCs.  For example, in a cylinder system $\psi|_{\partial M}=0$ only give a confined boundary condition in axial direction and there are periodic boundary condition around the axis in the cylinder surface.  A persistent SC would circuit around the axis in the cylindrical surface, see Fig.\ref{cylinder}.  Since a finite width rings, the common mesoscopic systems, are not simple connected, a circular persistent SC would exist in them.  It has been shown that the persistent SC can be detect by measure the induced electric field\cite{Sun04} or mechanical torque\cite{Sonin07}.

\begin{figure}[!hbp]
\centering
\includegraphics{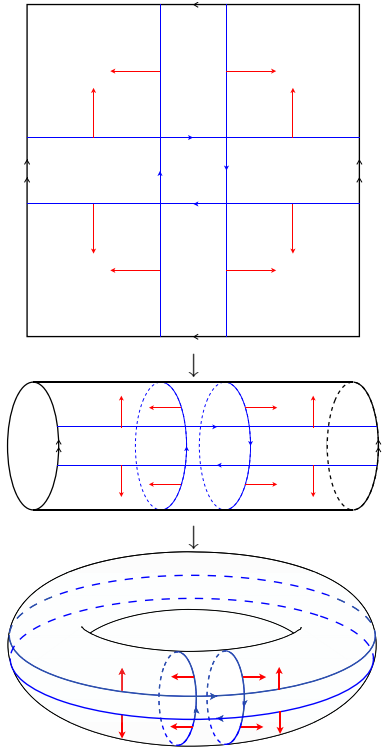}
\caption{Schematic diagram for the circular movement of spin in a torus. \label{torus}}
\end{figure}

 The SCs in the infinite 2D system found by Rashba is so mysterious that their existence and the conventional definition of the SC are doubted.  However, the persistent SC which is transport in a mesoscopic ring with SOI is satisfied the conventional definition of the SC.  Since the conventional definition of the SC makes physical sense, they might need not to be modified\cite{Sun08}.  It has been shown that in the finite 1DES and 2DES the nonzero persistent SCs which satisfy the conventional definition are all transport when the boundary conditions is considered.  It could be shown that the discrete wave functions considered by Rashba could exist in not only the infinite 2D topological space $\mathbb{R}^2$ but also the finite topological space of a torus $T^2$.  Since the wave functions is periodic in two dimensions, there are an equivalence relation $\sim$ which satisfies $(x_1,y_1)\sim(x_2,y_2)$ if $x_2=x_1+2\pi n_x$ and $y_2=y_1+2\pi n_y$.  The topological space of the system can also be the quotient space $\mathbb{R}^2/\sim$ which is the torus $T^2=S^1\times S^1$\cite{Nakahara90}. The first homotopy group $\pi_1(T^2)$ of a torus is isomorphic to $\mathbb{Z}\times\mathbb{Z}$\cite{Munkres00}.  Thus there might be two independent persistent SCs in the 2D torus SOI system, see Fig.\ref{torus}.  The background SCs in the infinite 2DES with SOI found by Rashba could be realized by the transport persistent SCs in the topological space of a torus.  In this sense, we meet the challenge proposed by Rashba.  

In summary, we investigate the persistent SCs in 1DES and 2DES with SOI in thermodynamic equilibrium at absolute zero temperature.  The persistent SC is the intrinsic one which is connected with Berry phase in the configuration space of a system and WNs of field configuration of electrons.  The existence of persistent SCs is only connected with the property of the first homotopy group of the configuration of a system.  The persistent SC would not be destroyed by impurity scattering in ballistic limit.  The background nonzero SCs found by Rashba in infinite 2DES with SOI could be realized by the transport persistent SCs in a finite torus electron system.  In this sense, we meet Rashba's challenge.

\begin{acknowledgments}
We thank Q. Q. Wang, N. N. Huang, Hugo de Garis, C. Hao, M. T. Cheng, Z. K. Guo and K. Zhu for helpful discussions.  This work is supported by NSFC under grant number 10534030. X. Z. acknowledges support from the China Scholarship Council.
\end{acknowledgments}


\begin{thebibliography}{}
\bibitem{tpcomment}
In literature,  the path $\theta$ is usually called field configuration.
\end{thebibliography}

\end{document}